\documentclass[11pt]{article}

\usepackage[sc,osf]{mathpazo}
\usepackage[top=1.5cm, left=2.5cm, right=3.0cm, bottom=2.0cm]{geometry}
\usepackage{graphicx, color}
\usepackage[dvipsnames]{xcolor}
\usepackage{amsmath, amssymb, amsfonts, amsthm, paralist}
\allowdisplaybreaks 
\usepackage{setspace}
\usepackage{url}
\usepackage[colorlinks,citecolor=blue,urlcolor=blue,linkcolor=blue]{hyperref}
\usepackage{bm}
\usepackage{authblk}
\usepackage{booktabs}
\usepackage{threeparttable}

\usepackage{mathtools}
\usepackage[round]{natbib}
\usepackage{paralist}
\usepackage{palatino}
\usepackage{courier}

\newcommand{\pr}{\mathrm{Pr}}
\newcommand{\T}{{\mathsf{T}}} 

\title{\textbf{Tractable Ridge Regression for Paired Comparisons}}
\author[1]{Cristiano Varin}
\author[2]{David Firth}
\affil[1]{Ca' Foscari University of Venice, Italy}
\affil[2]{University of Warwick, Coventry, UK}
\date{}

\begin{document}

\maketitle

\begin{abstract}
Paired comparison models, such as Bradley-Terry and Thurstone-Mosteller, are commonly used to estimate relative strengths of pairwise compared items in tour\-na\-ment-style data.   We discuss estimation of paired comparison models with a ridge penalty.  A new approach is derived which combines empirical Bayes and composite likelihoods without any need to re-fit the model, as a convenient alternative to cross-validation of the ridge tuning parameter.  Simulation studies demonstrate much better predictive accuracy of the new approach relative to ordinary maximum likelihood.  A widely used alternative, the application of a standard bias-reducing penalty, is also found to improve appreciably the performance of maximum likelihood; but the ridge penalty, with tuning as developed here, yields greater accuracy still.  The methodology is illustrated through application to 28 seasons of English Premier League football.

\textbf{Keywords:} Bradley-Terry model; Composite likelihood; Pairwise empirical Bayes; Probit regression; Rating; Shrinkage; Sport tournaments; Thurstone-Mosteller model.
\end{abstract}

\section{Introduction}\label{sect:intro}

The problem of rating a list of items on the basis of a set of paired comparisons arises frequently, in a variety of fields including artificial intelligence \citep[e.g.,][]{Rafailov:23}, bibliometrics \citep[e.g.,][]{Varin:16}, education \citep[e.g.,][]{Bartholomew:22}, forensic science \citep[e.g.,][]{Thompson:18}, genetics \citep[e.g.,][]{Ma:12}, politics \citep[e.g.,][]{Loewen:12}, psychometrics \citep[e.g.,][]{Maydeu-Olivares:05} and sport \citep[e.g.,][]{Glickman:17}, to name just a few. A classical reference for the statistical analysis of paired comparisons is \cite{David:88}. A more recent review is \cite{Cattelan:12}, while \cite{Aldous:17} discusses various aspects of paired comparisons from the point of view of applied probability. 
  
In binary paired comparison models, the probability that item $i$ beats item $j$ in a comparison between them is
\begin{equation}\label{eq:thurstone}
\text{Pr}(i \; \text{beats} \; j) = F(\mu_i-\mu_j), 
\end{equation}
where $F(\cdot)$ is the cumulative distribution function of a zero-symmetric continuous random variable and $\mu_i$ is the strength of item $i$\null. Popular choices for $F$ are the logistic distribution which gives the Bradley-Terry model \citep{Bradley:52} and the normal distribution which corresponds to the Thurstone-Mosteller model \citep{Thurstone:27, Mosteller:51}. These two models are quite similar in practice, given the well-known correspondence between the logit link assumed in the Bradley-Terry model and the probit link of the Thurstone-Mosteller model \citep[e.g.,][pages 246--247]{Agresti:02}.

Consider $p$ items. Maximum likelihood estimation of the vector of strength parameters $\bm{\mu}=(\mu_1, \ldots, \mu_p)^\T$ is well known to be problematic, in part because the estimate of $\mu_i$ diverges if item $i$ wins all of its comparisons or loses them all \citep[e.g.,][]{Kosmidis:21b}. Bias-reduced maximum likelihood \citep{Firth:93} overcomes such limitations of maximum likelihood, guaranteeing finiteness and also better frequentist properties of the strength estimates \citep{Kosmidis:21b}.  Finite estimates can also be obtained through other forms of penalized maximum likelihood \citep{Mease:03} or  Bayesian methods \citep{Caron:12}.

The present paper develops a highly tractable alternative regularization of the likelihood for the Thurstone-Mosteller model, using a ridge penalty that is tuned by using a new `pairwise empirical Bayes' method.  The predictive performance of this new approach is studied in simulation experiments and through application to match results from incomplete seasons of English Premier League football.

\section{Ridge Regression for Paired Comparisons}\label{sect:ridge}

We begin by assuming that the outcome of a paired comparison does not depend on the order in which the items are presented, which in sport applications means that there is no home-field effect; and we assume for now also that there are no ties. These assumptions will be relaxed later, in Section \ref{sect:home_ties}.

Denote the outcome of the paired comparison between items $i$ and $j$ as $Y_{ij}=1$ if $i$ wins and $Y_{ij}=-1$ if $j$ wins. 
In this paper we investigate a natural form of regularization for estimation of the strength parameters, based on the ridge-penalized log likelihood
\begin{equation}\label{eq:ridge}
\ell_\lambda( \bm{\mu})=\ell( \bm{\mu})- \frac{\lambda}{2} \sum_{i=1}^p \mu_i^2,
\end{equation}
where $\ell( \bm{\mu})$ is the log-likelihood of the paired comparison model, \emph{i.e.},
\begin{equation}\label{eq:Thurstone}
\ell(\bm{\mu}) = \sum_{(i,j) \in \mathcal S} \log F\{y_{ij}(\mu_i-\mu_j)\}.
\end{equation}
Here $\mathcal S$ denotes the observed tournament, \emph{i.e.}, $\mathcal S$ is the set of index pairs $(i, j)$ in which $y_{ij}$ is observed; and $\lambda\geq 0$ is a scalar tuning parameter. For a fixed value of $\lambda$, computation of the ridge estimates of the strength parameters is a relatively simple optimization, by using for example standard quasi-Newton algorithms. The standard approach to determining a good value for $\lambda$ is cross-validation.  An important point is that in pair-comparison data there is often some balancing structure that should be respected in cross-validatory splitting of the dataset; in round-robin tournaments, for example, it will usually make sense to use whole tournament rounds as cross-validation units.  When such structure is absent, though, it can often be difficult to identify a suitable set of splits for cross-validation with paired-comparison data. 

An alternative to cross-validation is estimation of $\lambda$ using the empirical Bayes method \citep{Morris:83}. Ridge regression coincides with maximum-a-posteriori estimation when the strength parameters are uncorrelated zero-mean normal random variables with precision $\lambda$,
\begin{equation}\label{eq:normal-strength}
 \bm{\mu}  \sim \mathcal N( \bm{0}, \lambda^{-1}  {\text I}_p),
\end{equation}
where $ {\text I}_p$ is the identity matrix of size $p$. Normality of the strengths is purely a working assumption here, to allow determination of $\lambda$ via the empirical Bayes method. In the simulation studies reported in Section \ref{subsect:simulations_t} we show that this method of determining $\lambda$ is robust in the sense that it works well even when the true strengths have appreciably heavier tails than a normal distribution.
(There is no need to consider violations of normality due to skewness, because the paired comparison model is identified by differences $\mu_i-\mu_j$, which have a symmetric distribution in any case.) 

The empirical Bayes method determines $\lambda$ by maximizing the marginal likelihood obtained
from integrating out the team strengths. This leads to a cumbersome integral of dimension $p$,
\begin{equation}\label{eq:likelihood}
L(\lambda)= \lambda^{p/2}  \int_{\mathbb{R}^p} \exp\{\ell(\bm{\mu})\} \prod_{i=1}^p   \phi( \lambda^{1/2}  \mu_i) \text d  {\mu},
\end{equation}
where $\phi(\cdot)$ is the density of the standard normal random variable.

\section{Pairwise Empirical Bayes}\label{sect:pairwise}

\subsection{Estimation of the tuning parameter}

We develop here a simple method that avoids the potentially high-dimensional integration (\ref{eq:likelihood}) by using composite likelihood methods \citep{Varin:11}. The tuning parameter $\lambda$ is estimated by maximizing the pairwise log-likelihood constructed from all couples of correlated paired comparisons, \textit{i.e.}, those couples with one item in common, such as ($i$ vs $j$) and ($i$ vs $k$). The pairwise log-likelihood for $\lambda$ takes the form 
\begin{equation}\label{eq:logpair}
\ell_{\text{pair}}(\lambda)= \sum_{r, s \in \mathcal O} n_{rs} \log p_{rs}(\lambda). 
\end{equation}
Here $\mathcal O=\{-1, +1\}$ is the set of possible outcomes of the paired comparisons (which will be later extended in Section \ref{sect:home_ties} to include ties); while $n_{rs}=\sum_{ijk} 1(Y_{ij}=r, Y_{ik}=s)$ is the number of correlated couples of paired comparisons with outcomes $r$ and $s$,
and $p_{rs}(\lambda)=\pr\left(Y_{ij}=r, Y_{ik}=s\right)$ is the corresponding marginal bivariate probability.

In the following we use the Thurstone-Mosteller model, because it is mathematically more convenient than the Bradley-Terry model and the two models give essentially equivalent fits as already noted in Section \ref{sect:intro}. Under the Thurstone-Mosteller model, the bivariate probabilities $p_{rs}(\lambda)$ have closed-form expressions, by a well-known result sometimes referred to as the Theorem of Median Dichotomy \citep{Sheppard:98, Cox:02}. The probability of two wins for the common item $i$ is 
\begin{align*}
p_{11}(\lambda)&=\Pr(Y_{ij}=1, Y_{ik}=1)\\
&=\int_{0}^{\infty} \int_{0}^{\infty} \phi_2\left(u, v; \frac{\lambda^{-1}}{1+2\lambda^{-1}}\right) \text d u \text d v\\
&=\frac{(1+\tau)}{4},
\end{align*}
where $\phi_2(u, v; \rho)$ is the density function of a bivariate standard normal variable with correlation $\rho$, and the Kendall rank correlation
\begin{equation}\label{eq:tau}
\tau=\frac{2}{\pi} \text{arcsin} \left(\frac{\lambda^{-1}}{1+2\lambda^{-1}}\right)
\end{equation}
takes a value in the interval $(0, 1/3)$. 
The general expression for $p_{rs}$ depends on whether there is concordance or not in the outcomes of the two paired comparisons:
$$
p_{rs}(\lambda)=
\begin{cases}
\displaystyle{\frac{(1+\tau)}{4}}, & \text{if} \; r=s,\\
\displaystyle{\frac{(1-\tau)}{4}}, & \text{if} \; r\neq s.
\end{cases}
$$
A closed-form estimate of $\lambda$  is obtained via reparameterization of the pairwise likelihood in $\tau,$ 
$$
\ell_{\text{pair}}(\tau)= c \log (1+\tau)+ d \log(1-\tau),
$$
where $c$ is the count of concordant correlated couples of paired comparisons ($Y_{ij}=Y_{ik}$) and $d$ the count of discordant couples ($Y_{ij}\neq Y_{ik}$).  The maximum pairwise likelihood estimator of $\tau$ is then the familiar Kendall $\tau$ coefficient
$$
\hat \tau=\frac{c-d}{c+d},
$$
and inversion of equation (\ref{eq:tau})  yields the maximum pairwise likelihood estimate of $\lambda,$  
\begin{equation}\label{eq:tau2lambda}
\hat \lambda=\frac{1-2\sin( \hat \tau \pi /2 )}{\sin( \hat \tau \pi/2  )}.
\end{equation}

In sparse datasets with only a modest number of comparisons per item, it is helpful to consider making a small-sample adjustment along the lines of the classical `rule of succession' of \citet{Laplace:14}.  In the present context such an adjustment adds one imaginary concordant couple and one imaginary discordant couple per item, and the corresponding small-sample adjusted estimator of $\tau$ is
$$
\hat \tau=\frac{c-d}{c+d+2p}.
$$
An alternative interpretation of this adjusted estimator is that it maximizes a penalized version of the pairwise log-likelihood, namely 
\begin{equation}\label{eq:small-sample}
\tilde{\ell}_{\text{pair}}(\tau)=  {\ell}_{\text{pair}}(\tau) + p \log(1-\tau^2).
\end{equation}

\subsection{Order effect and ties}\label{sect:home_ties}
The order of presentation of the items is important in many paired comparison contexts, notable examples being the home-field effect in sport competitions or the white player advantage in chess. In paired-comparison models, such an order effect is typically described through the inclusion of an intercept term (\emph{e.g.}, \citeauthor{Agresti:02}, \citeyear{Agresti:02}), 
\begin{equation}\label{eq:order-effect}
\Pr(Y_{ij}=1\mid\bm{\mu})=\Phi(\delta + \mu_i - \mu_j),
\end{equation}
where $\Phi(\cdot)$ is the distribution function of a standard normal variable.

Another common complication in paired comparisons is the presence of ties, represented by the null outcome $Y_{ij} = 0$ whenever $i$ and $j$ are compared but neither beats the other. 
As discussed in \cite{Agresti:92}, ties can be incorporated in paired comparison models through a cumulative link model. The Thurstone-Mosteller model with an order effect corresponds to the latent continuous process
$Z_{ij}=\delta+\mu_i-\mu_j + \epsilon_{ij}, $
where $\epsilon_{ij}$ are uncorrelated standard normal variables. The outcome of the comparison depends on a threshold parameter $\gamma \geq 0$, so that $j$ beats $i$ if $Z_{ij}< -\gamma$, $i$ ties with $j$ if $-\gamma \leq Z_{ij} < \gamma$ and $i$ beats $j$ if $Z_{ij} \geq \gamma.$ The corresponding cumulative probabilities for the three outcomes given the strengths are
$$
\Pr(Y_{ij} \leq x \mid \bm{\mu})=\Phi(c_x -\mu_i+\mu_j), \quad x \in \{-1, 0, 1\},
$$
for cutpoints $-\infty=c_{-2} < c_{-1} \leq c_0 < c_1=+\infty,$ with $c_{-1}=-\gamma-\delta$ and $c_0=\gamma-\delta$.
Ridge estimates of the strength parameters are still computed by maximizing a penalized likelihood as in (\ref{eq:ridge}), but now with the log-likelihood 
$$
\ell(\bm{\mu})=\sum_{(i,j) \in \mathcal S} \log\{ \Phi(c_{y_{ij}} -  \mu_i + \mu_j) - \Phi(c_{y_{ij}-1} - \mu_i + \mu_j) \}.
$$
As before, we proceed by empirical Bayes estimation of the model parameters $\gamma$, $\delta$ and $\lambda$ from the marginal distribution of the paired outcomes, and then compute the ridge estimates of the strength parameters. The marginal probabilities of the three possible outcomes are simply calculated by setting all strength parameters to zero, giving $\Pr(Y_{ij}=x)=\Phi(c_x)-\Phi(c_{x-1})$, for $x \in \{-1, 0, 1\}$. From these marginal probabilities, consistent estimates of $\gamma$ and $\delta$ are readily obtained as 
\begin{align}\label{eq:gamma}
\hat \gamma&=\frac{1}{2}\left\{\Phi^{-1}(1-\hat p_{1})-\Phi^{-1}(\hat p_{-1})\right\},  \\  \label{eq:delta}
\hat \delta&=\frac{1}{2}\left\{\Phi^{-1}(\hat p_{1})-\Phi^{-1}(\hat p_{-1}) \right\},
\end{align}
where $\hat p_{-1}=(n+1)^{-1}\sum_{ij} 1(Y_{ij}=-1)$ and $\hat p_1=(n+1)^{-1}\sum_{ij} 1(Y_{ij}=1)$. In tournaments without ties, $\hat \gamma=0$ as expected. 

Having estimated $\gamma$ and $\delta$, we proceed with estimation of the tuning parameter $\lambda$ by maximizing the pairwise log-likelihood (\ref{eq:logpair}), where $p_{rs}(\lambda)$ now denotes the joint probability of the outcomes of two paired comparisons with the same item appearing in either the first or the second position. For example, the probability that the first item $i$ obtains a win and a tie in such a couple of paired comparisons is 
\begin{align*}
p_{10}(\lambda) &= \Pr(Y_{ij}=1, Y_{ik}=0)  \\ 
&=\int_{c_0}^{c_1} \int_{c_{-1}}^{c_0} \phi_2\left(u, v; \frac{\lambda^{-1}}{1+2\lambda^{-1}}\right) \text d u \text d v \\
&=\int_{\gamma-\delta}^{\infty} \int_{-\gamma-\delta}^{\gamma-\delta} \phi_2\left(u, v; \frac{\lambda^{-1}}{1+2\lambda^{-1}}\right) \text d u \text d v.
\end{align*}
The other bivariate probabilities involved in the pairwise likelihood are similarly computed. In these probabilities $\gamma$ and $\delta$ are replaced with their estimates (\ref{eq:gamma})-(\ref{eq:delta}) so that the pairwise likelihood depends on $\lambda$ only. 

Maximization of the pairwise likelihood for $\lambda$ is conveniently performed by using the robust algorithm of \citet[ch.~4]{Brent:13} on the finite interval domain of $\tau$ from equation (\ref{eq:tau}), with subsequent back-transformation as in equation (\ref{eq:tau2lambda}). The same small-sample adjustment as before remains available through the pairwise likelihood penalty of equation (\ref{eq:small-sample}).

\section{Simulations}\label{sect:simulations}

\subsection{True strengths normally distributed}

The  performance of the pairwise empirical Bayes method is illustrated first by simulations from the Thurstone-Mosteller model with normally distributed strength parameters. We considered 120 scenarios given by the combination of: three values for the number of items $p\in\{20, 40, 60\}$; eight values for the tuning parameter $\lambda \in \{2^{-1}, 2^0, \ldots, \allowbreak 2^6\}$; and five sample sizes corresponding to five increasing fractions of the paired comparisons in a double round-robin tournament. More specifically, we simulated the full double-round-robin tournament and used the first $m$ rounds for training and the remaining $2(p-1)-m$ rounds for testing. The number of training rounds was chosen so as to approximate a specified proportion of the tournament paired comparisons. For example, when we consider 20\% of paired comparisons for training and the number of items is $p=30$, the training set is taken to be the first 12 rounds because $0.2\times (2\times 29)=11.6$\null. 
The five sample fractions considered for training in our simulations are $20\%, 30\%, 40\%, 50\%$ and $80\%$ of the paired comparisons in the double round-robin tournament.  For each scenario, $1000$ replications were generated and analysed. 

\begin{figure*}[t] 
\centerline{\includegraphics[scale=.6]{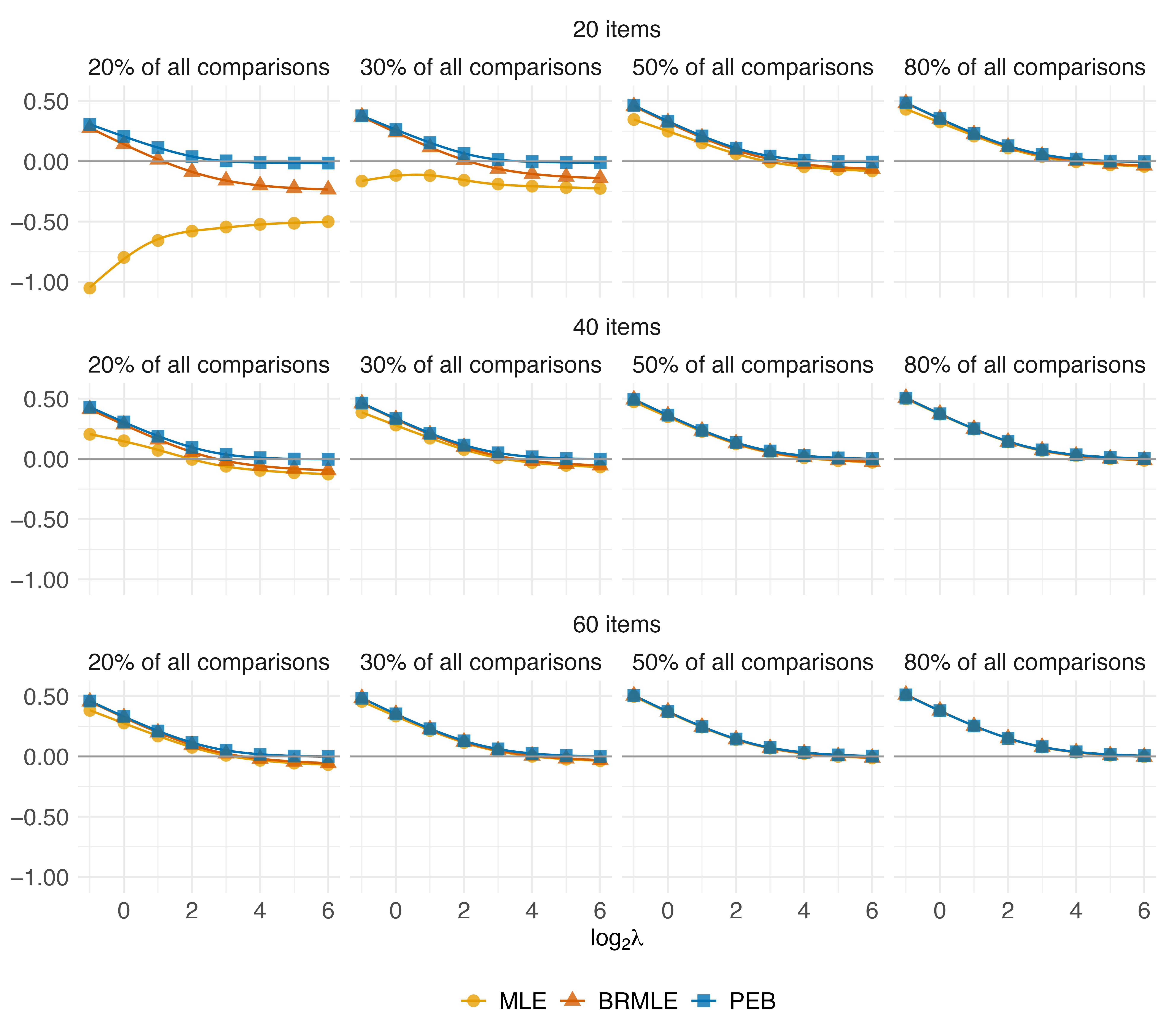}}
\caption{Logarithmic skill scores computed for incomplete double round-robin tournaments of different sizes. The columns correspond to various levels of incompleteness (20\%, 30\%, 50\% or 80\% of all possible paired comparisons used for training and the rest of paired comparisons for prediction). The rows correspond to different numbers of items (20, 40, 60). The item strengths are drawn from a zero-mean normal distribution with precision $\lambda$. The strengths are estimated by bias reduced maximum likelihood (BRMLE), maximum likelihood (MLE) and pairwise empirical Bayes (PEB). The reference method is a naive forecast that uses only the assumed marginal probability of a win for the first item and corresponds to a zero logarithmic skill score (grey solid line).}\label{fig:simul_normal}
\end{figure*}

The performance of our pairwise empirical Bayes method is compared here with three other methods: (1) maximum likelihood estimation, (2) bias-reduced maximum likelihood as in \citet{Firth:93}, and (3) maximum ridge-penalized likelihood with $\lambda$ determined by cross-validation.  Bias-reduced maximum likelihood estimation is computed through the \textbf{R} \citep{R:22} package \textbf{brglm2} \citep{Kosmidis:21a}. Since paired-comparison models are specified through pairwise differences of strengths, an arbitrary constraint is needed for identification of the maximum likelihood and bias-reduced maximum likelihood estimates; in the simulations we used the constraint $\mu_1=0$.  Cross-validated ridge estimates were calculated via the \textbf{glmnet} package \citep{Friedman:10, Tay:23}, using tournament rounds as the assessment unit. Paraphrasing the typical language used in model validation, we could call such a design a `leave one round out' cross-validation.

We simulated paired comparisons without ties, in order to focus on the estimation of the strength parameters. Estimation of the tie threshold parameter $\gamma$ would not be problematic, but bias-reduced maximum likelihood and cross-validated logistic regression are computationally less expensive (for repeated use in a simulation experiment) in paired comparison models without ties. 
In our simulations the order effect parameter was set to $\delta=0.2$, meaning that in approximately 58\% of paired comparisons the first item beats the second (since $\Phi(0.2)\approx 0.58$).

The predictive performance of each method is measured by the classical logarithmic score (LS), defined as the negative average log-likelihood for the testing paired comparisons computed at the strengths estimated from the training comparisons. As reference we consider the naive forecast that predicts the result of every paired comparison using the assumed  marginal probability of a win for the first item, that is $\text{LS}_{\text{naive}}=-0.58 \log(0.58)-0.42 \log(0.42)=0.68$. The naive forecast corresponds to predictions for the paired comparison model with strength parameters all zero, $\mu_i=0$ for all $i=1, \ldots, p$.
With this reference, we define the logarithmic skill score as
$
\text{LSS}=1-\text{LS} /\text{LS}_{\text{naive}}=1-\text{LS} / 0.68.
$
The higher the logarithmic skill score, the better the predictive performance. Better forecasts on average than those obtained by the naive forecast have a logarithmic skill score above zero.

The simulation results are summarized in Figure \ref{fig:simul_normal} which displays the average of the simulated logarithmic skill scores. 
We do not report results based on cross-validation in Figure \ref{fig:simul_normal}, because they were almost everywhere visually indistinguishable from those based on the adjusted pairwise empirical Bayes method. 
This finding confirms that pairwise empirical Bayes for the Thurstone-Mosteller model is essentially equivalent to standard cross-validation, but with the substantial computational advantage that no repeated refitting of the model is needed.

We now comment on the results reported in Figure \ref{fig:simul_normal}. As expected, as the number of items or the fraction of paired comparisons used for training increases, all methods converge to the same predictive performance. Furthermore, as $\lambda$ increases, estimates of item strengths converge to zero and therefore all methods produce predictions that are on average equivalent to those of the naive forecast. The predictive performance of maximum likelihood estimation is completely unsatisfactory when the training set is small and the number of items is 20, with its average logarithmic skill scores being worse than the naive forecast. 
The bias-reduced maximum likelihood estimator improves substantially on maximum likelihood in most settings, but with small training sets and small values of $\lambda$ it still leads to worse logarithmic skill scores on average than the naive forecast.

As shown in Figure \ref{fig:simul_normal}, pairwise empirical Bayes has uniformly the best predictive performance of all methods considered.  Pairwise empirical Bayes was even found to perform relatively well in a still more extreme setting, with $p=20$ items and only 4 paired comparisons per item (approximately 10\% of the tournament) used for training; that extreme setting is not displayed in Figure \ref{fig:simul_normal} because maximum likelihood performs so poorly that its logarithmic skill score values fall too far below zero. 

\subsection{True strengths distributed as Student t}\label{subsect:simulations_t}

The results just described were obtained under the assumption of normally distributed strengths, the assumption used to construct the pairwise empirical Bayes method. Less benign settings for pairwise empirical Bayes are considered next, by drawing the strengths from Student's $t$ distribution with $\nu$ degrees of freedom. The simulated strengths are scaled so that their precision continues to be $\lambda$, 
$\mu_i  \sim t_\nu \sqrt{(\nu - 2)/(\lambda\nu)}$, $\nu > 2$, $i=1, \ldots, p,$ 
with $t_\nu$ denoting a Student's $t$-distributed variate with $\nu$ degrees of freedom. We take $\nu > 2$ since otherwise the variance would be undefined. Drawing the strengths from a $t$ distribution is of interest because it allows a small number of items to be distinctly stronger or weaker than the rest. Figures S1 and S2 of the Supplementary Material display the average logarithmic scores for the same 120 scenarios considered previously, but with strengths simulated from Student's $t$ distributions with $\nu=8$ and $\nu=3$ degrees of freedom. As far as predictive performance is concerned, there is no perceptible difference between the $t_8$ and normal distributions; and even in the more extreme scenario with $t_3$-distributed strengths, the differences are rather small.  While such simulation evidence is inevitably limited, it does provide some reassurance that the pairwise empirical Bayes method is robust to excess kurtosis in the strengths. 

\section{Premier League}\label{sect:application}

As an illustration of the ridge method we consider estimation of the strength of football teams in a round-robin tournament, before the tournament is complete.
Since season 1995--1996, the English Premier League has been a double round robin between twenty teams. The current format of the tournament therefore consists of 38 match-weeks.  We consider here all the results of the $10,640$ matches of the 28 seasons between 1995 and 2023, in order to evaluate the robustness of our method over a long period characterized by substantial changes in (English) football. For each season, we trained the paired comparison models with the first 10, 15, 20, 25 and 30 match-weeks and predicted the results of the rest of the tournament. 

The problem we consider in this illustration is different from sequential prediction of next week's match results given the team's past results, for which it would be more appropriate to consider dynamic paired comparison models, such as the state-space approaches of \cite{Knorr:00, Fahrmeir:94}, the exponential smoothing of \cite{Cattelan:13}, the time-weighted likelihood of \cite{Mchale:11}, the kernel smoothing of \cite{Bong:20}, or the recent nonparametric spectral ranker of \cite{Tian:24}.

In our illustration, the Thurstone-Mosteller model with ties (which here will be called `draws', the word used in football) was estimated by maximum likelihood estimation using the \textbf{polr} function from the \textbf{MASS} package \citep{MASS:02}, by bias-reduced maximum likelihood for the cumulative logistic regression model developed in \cite{Kosmidis:14} and implemented in the R function \textbf{bpolr} available through the supplementary materials of that paper, and by our pairwise empirical Bayes method using R code available at  \url{https://github.com/crisvarin/peb}. 

As a reference forecast, we considered the naive predictions calculated from long-term frequencies of home wins, draws and away wins pooled across all 28 seasons of the Premier League. Those seasons saw 46\% home wins, 25\% draws and 29\% away wins out of a total number of $10,640$ matches. The logarithmic score for the naive forecast is thus
$\text{LS}_{\text{naive}}=-0.46\log(0.46)-0.25\log(0.25)-0.29\log(0.29)=1.06.$
 
\begin{figure*}[!h]
\centerline{\includegraphics[scale=.65]{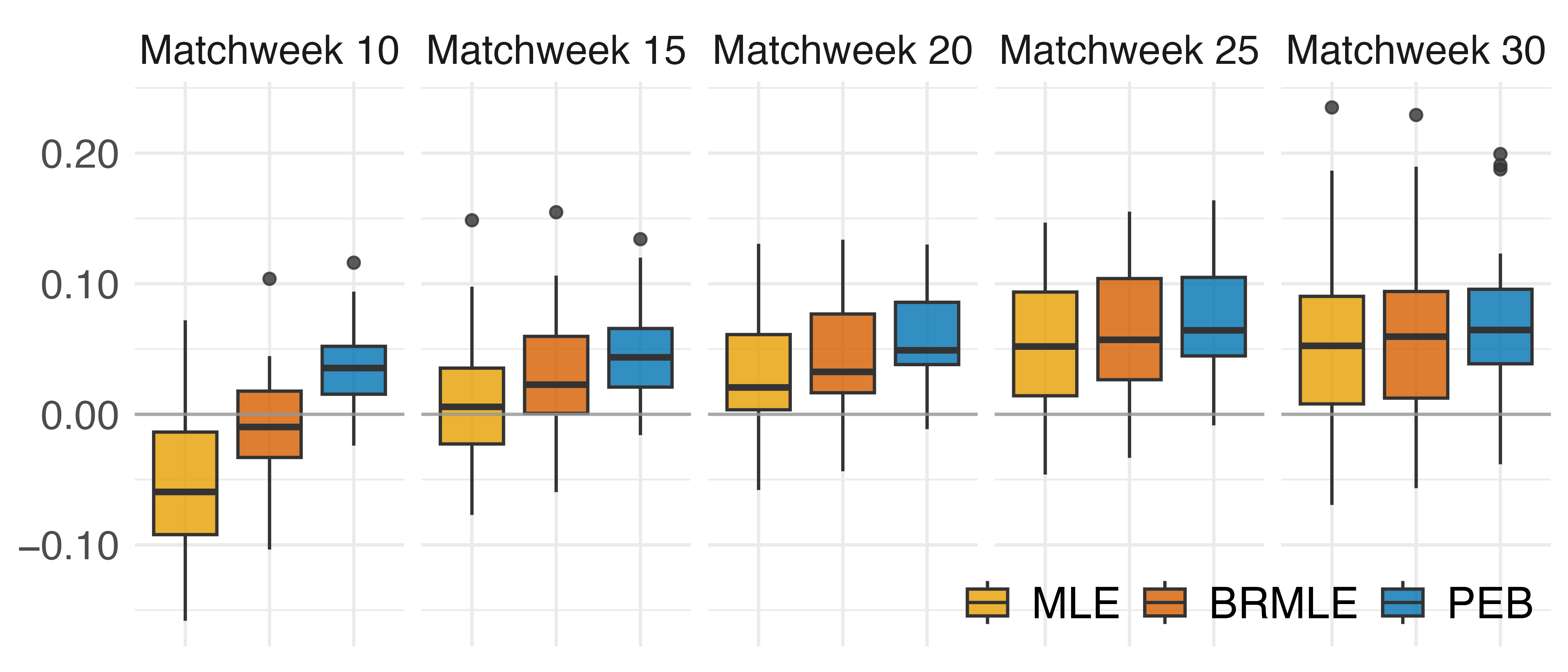}}
\caption{Boxplot summary of the distributions, across 28 Premier League seasons, of the logarithmic skill scores computed after 10, 15, 20, 25 and 30 matchweeks.  The predictions are computed with strengths estimated using maximum likelihood (MLE), bias-reduced maximum likelihood (BRMLE) and pairwise empirical Bayes (PEB). The reference method is a naive forecast that uses only long-term frequencies of home wins, draws and away wins and corresponds to a zero logarithmic skill score (grey solid line).}\label{fig:logscores_premier_aggregated}
\end{figure*}

Figure \ref{fig:logscores_premier_aggregated} displays boxplots of the collected logarithmic skill scores: each boxplot is calculated from the 28 logarithmic skill scores for the various seasons, in a given match-week. 
The boxplots show that the predictive performance of pairwise empirical Bayes is substantially better than that of maximum likelihood estimation, in particular when the training data are match results from only the first 10 or 15 weeks of the season. At those early stages of the tournament, just as we saw in the simulation studies of Section \ref{sect:simulations}, maximum likelihood very often produces strength-parameter estimates that predict future matches worse than does the naive forecast.  

The results here also show that the bias-reduction method of \citet{Firth:93} improves upon maximum likelihood estimation in terms of prediction: in all 28 seasons and for all sizes of the training data, the logarithmic skill scores based on bias-reduced maximum likelihood are better than those of standard maximum likelihood. However, pairwise empirical Bayes typically performs even better than bias-reduced maximum likelihood. For prediction in the early weeks of the season, especially, pairwise empirical Bayes out-performs the other methods by a substantial margin. 

\section{Final Remarks}
The main purposes of this paper were to discuss ridge-penalized estimation of paired-comparison models, and to develop the simple and effective pairwise empirical Bayes method for tuning the ridge penalty in the Thurstone-Mosteller model.  As expected, the shrinkage provided by a ridge penalty yields predictions with substantially higher precision than standard maximum likelihood estimation. Indeed when a paired-comparison model is trained with relatively few observations, prediction based on maximum likelihood is actually worse than the naive predictor that uses only the marginal proportions of (home win, away win, draw) outcomes. The poor predictive performance of maximum likelihood when the training set is small is due to overfitting; the shrinkage achieved through a ridge penalty, tuned by the pairwise empirical Bayes method, effectively eliminates that problem.

The simulations of Section \ref{sect:simulations}  show that the amount of shrinkage needed for good predictive performance, especially when the training set and the number of items are both small, is appreciably greater than the shrinkage inherent in the bias-reduced maximum likelihood method of \cite{Firth:93}.  This is unsurprising: it is well known that unbiasedness typically is not the main consideration for predictive performance.

The authors are not aware of other work focused on estimation of paired comparison models with a ridge penalty. 
Some previous work in team sports modeling has considered more traditional linear ridge penalty models to assess the contribution of individual players to a team's performance. A recent example is the adjusted plus/minus regularized model discussed in \cite{matano:23}; see also the references therein.

\section*{Funding}
The work of both authors was supported by the IRIDE grant 2023 from DAIS, Ca' Foscari University.

\section*{Supplementary Material}
\label{SM}
Supplementary material provides the simulation results described in Section \ref{subsect:simulations_t} where strengths are drawn from a Student $t$ distribution. The Premier League data and R code to reproduce the illustration of Section \ref{sect:application} are available in the public repository \url{https://github.com/crisvarin/peb}.

\bibliographystyle{chicago} 
\bibliography{varin-firth-peb-arXiv.bib}

\end{document}